\documentclass[aps,prl,reprint]{revtex4-1}
\usepackage[margin=1in]{geometry}
\usepackage{amsmath}
\usepackage{amssymb}
\usepackage{graphicx}
\usepackage{wasysym}
\usepackage{braket}

\newcommand{\mass}{m}
\newcommand{\lspacing}{d}
\newcommand{\tparam}{t}
\newcommand{\uparam}{U}

\begin{document}
\title{Disappearance of Quasiparticles in a Bose Lattice Gas}
\author{David Chen}
\altaffiliation[now at:]{Max Planck Institute of Molecular Cell Biology and Genetics, Pfotenhauerstr. 108,01307, Dresden, Germany}
\affiliation{Department of Physics, University of Illinois at Urbana-Champaign, Urbana, IL 61801, USA}
\author{Carolyn Meldgin}
\affiliation{Department of Physics, University of Illinois at Urbana-Champaign, Urbana, IL 61801, USA}
\author{Philip Russ}
\affiliation{Department of Physics, University of Illinois at Urbana-Champaign, Urbana, IL 61801, USA}
\author{Erich Mueller}
\affiliation{Department of Physics, Cornell University, Ithaca, NY 14853, USA}
\author{Brian DeMarco}
\email{bdemarco@illinois.edu}
\affiliation{Department of Physics, University of Illinois at Urbana-Champaign, Urbana, IL 61801, USA}

\date{\today}

\begin{abstract}
We use a momentum-space hole-burning technique implemented via stimulated Raman transitions to measure the momentum relaxation time for a gas of bosonic atoms trapped in an optical lattice.  By changing the lattice potential depth, we observe a smooth crossover between relaxation times larger and smaller than the bandwidth.  The latter condition violates the Mott-Ioffe-Regal bound and indicates a breakdown of the quasi-particle picture.  We produce a simple kinetic model that quantitatively predicts these relaxation times.  Finally, we introduce a cooling technique based upon our hole-burning technique.
\end{abstract}

\maketitle

Numerous quantum many-particle systems, such as electrons in metals, are surprisingly well modeled by gases of weakly interacting quasiparticles.  This remarkable observation provides one of the most powerful paradigms in quantum many-body physics and underpins many calculational techniques.  A quasiparticle is typically viewed as a composite formed from an elementary particle (e.g., an electron) and a cloud of excitations (e.g., phonons or particle-hole pairs) \cite{MahanBook}.  The aggregate object behaves like a particle: it has a well-defined quasimomentum $k$, energy $\epsilon_k$, and mean-free path $\lambda_k$.  The latter is the average distance the quasiparticle travels before being scattered into another direction.  In 1960, Ioffe and Regel pointed out that this picture is sensible only if the mean-free path is larger than the lattice spacing in the material \cite{ioffe1960non}, an idea which was further developed by Mott in 1972 \cite{mott1972conduction}.  In solids, estimates of the mean-free path from resistivity indicate that most metals satisfy this Mott-Ioffe-Regel (MIR) bound.  Those materials that violate the MIR limit typically have other unusual properties \cite{RevModPhys.75.1085,hussey2004universality}.  For example, many have a resistivity that grows linearly with temperature $T$, in contrast to the $T^2$ prediction of Fermi-liquid theory.  An equivalent expression of the MIR bound is that, within a quasiparticle picture, the quasi-momentum relaxation time $\tau$ associated with $\lambda_k$ cannot be shorter than $h/W$, where $W$ is the bandwidth (i.e., the difference between the largest and smallest $\epsilon_k$), and $h=2\pi\hbar$ is Planck's constant.

As we demonstrate, the disappearance of quasiparticles need not be a sudden event.  For bosonic atoms trapped in an optical lattice, we find no sharp transition in the quasi-momentum relaxation time, which gradually decreases as the inter-particle interaction strength is increased.  For weak interactions (compared with tunneling) the relaxation is slow, indicating that the system is well described by quasiparticles.  In contrast, for strong interactions, relaxation is sufficiently fast to violate the MIR bound and suggest that the quasiparticle concept cannot be applied.  These regimes are continuously connected, and we cannot demarcate an abrupt boundary between them.  This observation is consistent with the idea that the MIR bound is not a sharp inequality, but rather a scale describing a crossover between behaviors.

We study a gas of $^{87}$Rb atoms trapped in a three-dimensional cubic optical lattice in the superfluid regime of the Bose-Hubbard model.  The atoms are cooled to the lowest energy band of the lattice in the $\ket{\downarrow}=\ket{F=1,m_F=0}$ hyperfine state.  Significant occupation of high quasimomentum states is present, because the temperature of the gas is comparable to the bandwidth $W=12t$, where $t$ is the tunneling energy.  As illustrated in Fig. 1(a), we use stimulated Raman transitions to drive the atoms to the first-excited band and $\ket{\uparrow}=\ket{F=2,m_F=0}$ state. Because the Raman wavevector difference $\Delta k=1.45 \pi/d$ lies along the $z$-direction of the lattice, the Raman transitions can only change the quasimomentum component $q_z$ by $\Delta q_z=2 q_B-\hbar\Delta k\approx 0.55 q_B$, where $q_B=\hbar \pi/d$ and $d=406$~nm is the lattice spacing.  The strong dispersion of the first-excited band enables us to tune the frequency difference $\Delta\omega$ between the Raman beams so that atoms with $q_z\approx 0$ are resonant \cite{Zoller2007}.  Fig. 1(b) shows images of quasimomentum distributions illustrating that these atoms are transferred to the $\ket{\uparrow}$ state.  We subsequently eject these atoms from the trap using a 50~$\mu$s pulse of light resonant with the $5S_{1/2},F=2\rightarrow 5P_{3/2}$ transition.

\begin{figure}[h!]
\includegraphics[width=0.9\columnwidth]{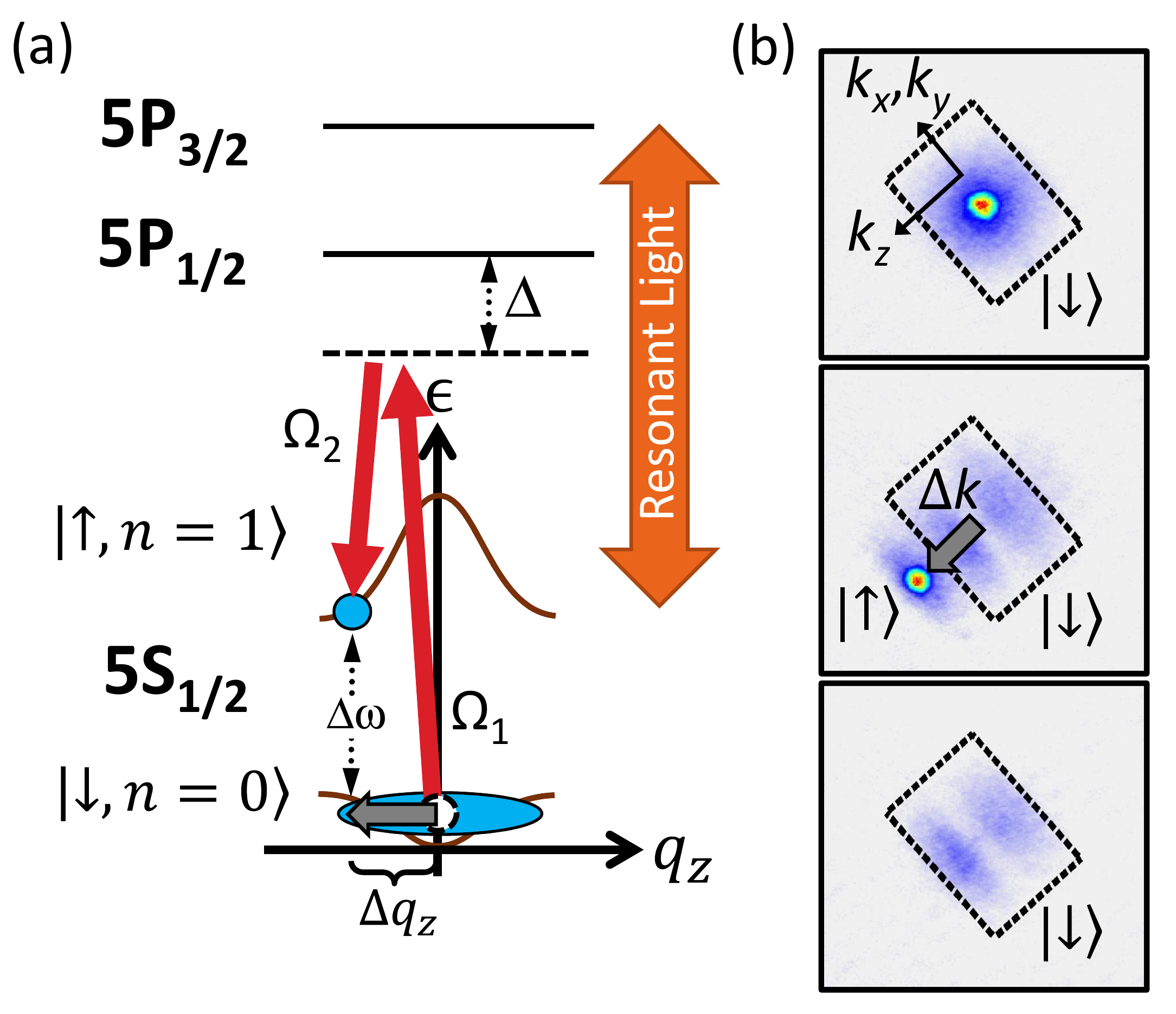}
\caption{\label{fig:RamanConfig}Quasimomentum-selective stimulated Raman transition and procedure for measuring quasimomentum relaxation. (a) Energy $\epsilon$ of $^{87}$Rb atoms confined in a lattice potential.  Only the ground band ($n=0$) in $\ket{\downarrow}$ and first-excited band ($n=1$) in $\ket{\uparrow}$ are shown. The Raman beams (red arrows), with Rabi rates $\Omega_1$ and $\Omega_2$ and detuned $\Delta$ from the transition to the $5P_{1/2}$ excited electronic state, are used to selectively transfer atoms between bands.  (b) Absorption images taken after bandmapping and 20 ms of TOF. The images are obtained before (top) and immediately after (middle) the Raman pulse and subsequent to the resonant-light pulse (bottom). The reciprocal lattice vectors are indicated by $k_x, k_y$, and $k_z$. The projection of the FBZ onto the imaging plane is displayed with dashed lines. Atoms in the first-excited band appear outside the FBZ after bandmapping.}
\end{figure}

\begin{figure}[h!]
\includegraphics[width=\columnwidth]{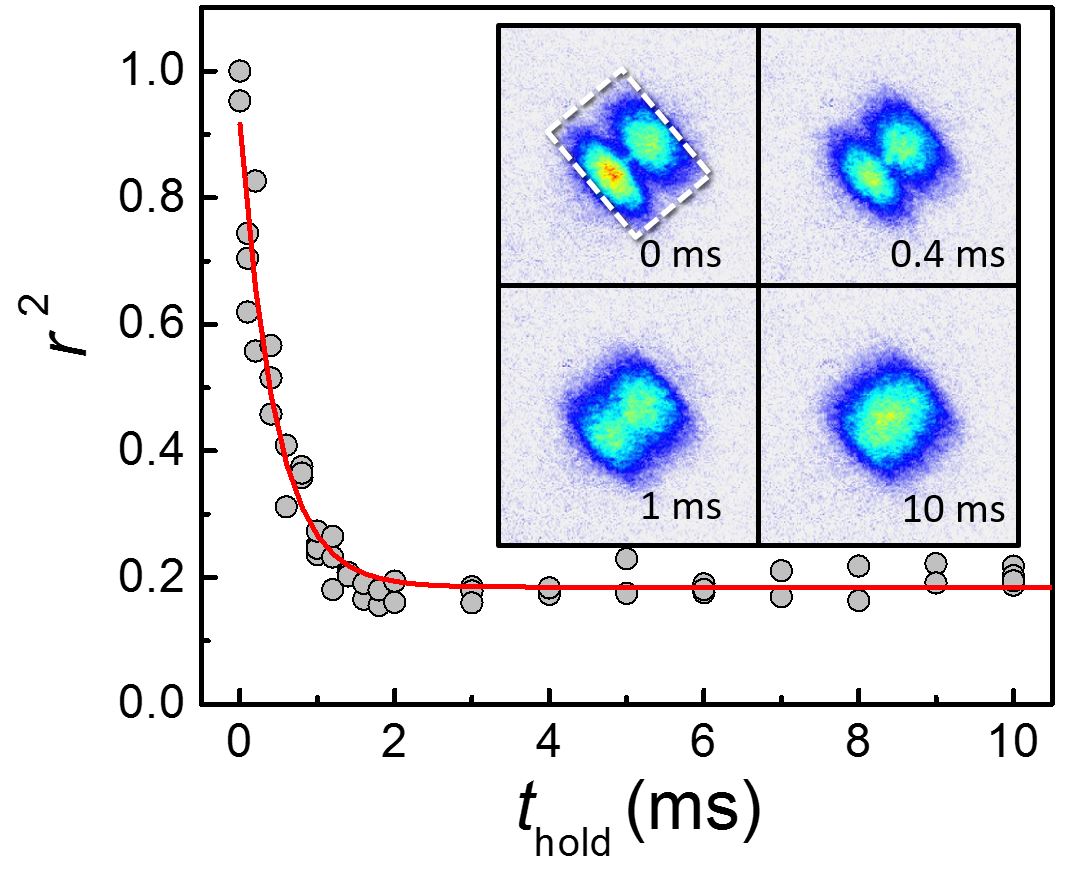}
\caption{\label{fig:r2vsholdtime}Sample relaxation data.  The mean squared residual of the quasimomentum-distribution fit $r^2$ versus hold time in the lattice $t_\text{hold}$ is shown for $s=6\,E_R$. Each data point corresponds to a single measurement, and the vertical axis of the plot has been re-scaled. The offset from $r^2=0$ is consistent with the residual from equilibrium images and is likely due to the failure of bandmapping at the edge of the FBZ (marked by the dashed white lines) and imaging noise. The red line is a fit to a single exponential decay used to determine the relaxation time constant $\tau$. }
\end{figure}

\begin{figure*}[ht!]
\includegraphics[width=2\columnwidth]{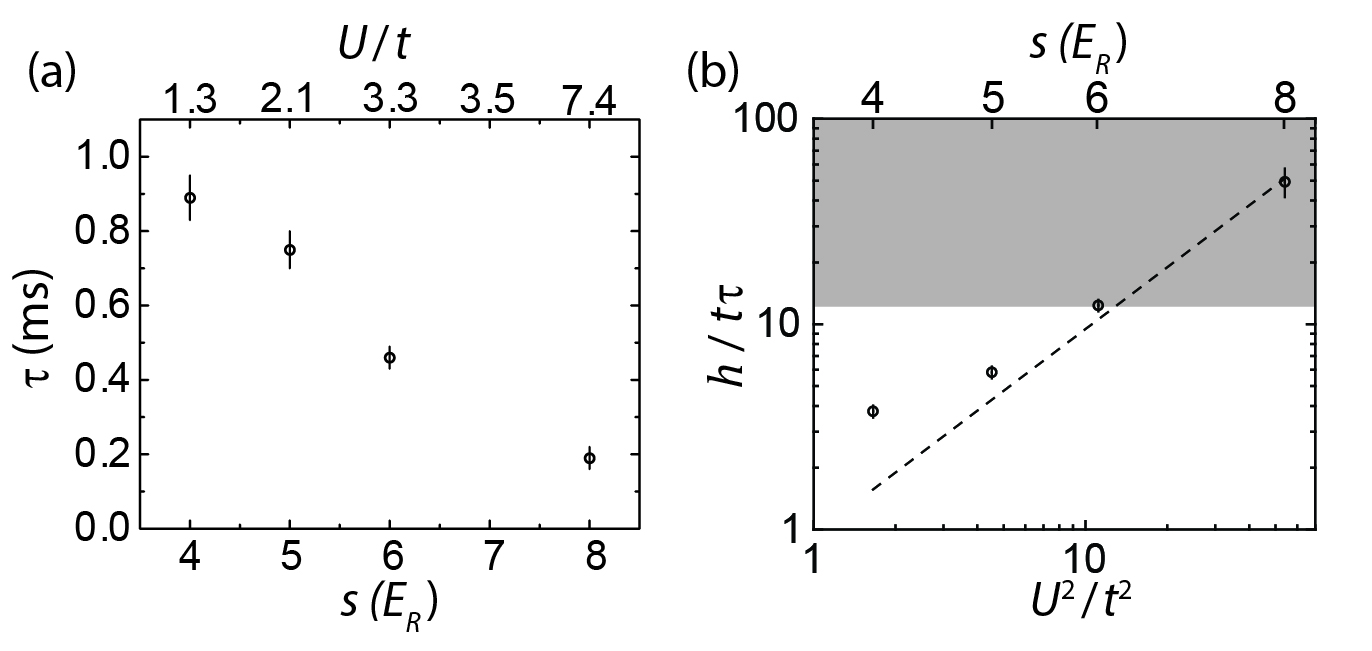}
\caption{\label{fig:tauvss} Quasimomentum relaxation measurements at different lattice potential depths.  (a) The measured relaxation time $\tau$ is shown for varied lattice potential depth $s$. Each point is determined using a fit to data such as those shown in Fig. 2, and the error bar displays the fit uncertainty. (b) Normalized relaxation rate $h/t\tau$ versus $U^2/t^2$. The dashed line represents the scaling law predicted by Eq.~\eqref{eq:a}. The error bars include the fit uncertainty used to determine $\tau$ and the standard deviation in measurements of $T$ and $N$. The MIR bound is violated in the region marked in gray, which corresponds to $h/\tau W>1$.}
\end{figure*}

After removing the low $q_z$ atoms, we allow the remaining atoms to evolve in the lattice potential for a time $t_\text{hold} = \left(0-10\right)$~ms.  We measure the quasi-momentum distribution after $t_\text{hold}$ using band-mapping and time-of-flight (TOF) imaging \cite{denschlag,McKay2009}.  As shown in the inset of Fig. 2, the quasi-momentum profile rapidly relaxes after the $q_z\approx0$ atoms are removed.  We use the mean squared residual of a fit to such images to determine the deviation from equilibrium at each $t_{\rm hold}$ and measure the relaxation timescale. The images are fit to a semi-classical model that describes the equilibrium quasimomentum distribution of a non-interacting bosonic gas trapped in a combined lattice--parabolic potential. The mean squared residual of the fit is $r^2 = \sum_{ij} \left(\text{OD}_{ij} - n_{ij}\right)^2/ \sum_{ij} 1$, where $\text{OD}_{ij}$ is the measured optical depth, $n_{ij}$ is the fit function \cite{McKay2009,SM}, and the summations are over the indices $i$ and $j$ within a mask defined by the first Brillouin zone (FBZ).  The residual $r^2$ is measured as $t_\text{hold}$ is varied at four different lattice depths: $s=4, 5, 6$ and $8\,E_R$, where $E_R = \hbar^2 \pi^2/2md$ is the recoil energy, and $m$ is the atomic mass.  We observe that the quasimomentum relaxation is exponential in time, as shown in Fig.~\ref{fig:r2vsholdtime} for the particular case of $s=6\,E_R$.  The measured relaxation time constant $\tau$ determined from a simple exponential decay for the four lattice depths is shown in Fig.~\ref{fig:tauvss}a.

The relaxation is comparable to or faster than the tunneling time $\hbar/\tparam = (0.5-1.5)$~ms and similar to the Hubbard interaction time $\hbar/U = (0.4-0.2)$~ms for $s=6$--8~$E_R$, where $U$ is the interaction energy for two atoms to occupy a site.  Furthermore, $\tau$ is approximately two orders of magnitude faster than the time constant observed in the absence of the lattice \cite{SM}. The relaxation process speeds up for a stronger lattice potential, ultimately reaching just 200~$\mu$s at $s=8$~$E_R$.

We compare the measured time constant to a short-range, two-body scattering calculation based on Fermi's golden rule (FGR), treating the Hubbard interaction term as a perturbation to the single-particle tight-binding Hamiltonian. We consider a thermal gas at equilibrium in a cubic lattice potential and calculate the rate of scattering into the $n_{\vec{q}=0}$ state. We identify the relaxation time $\tau$ with the inverse of this rate,
\begin{equation}
\label{eq:a}
\frac{1}{\tau} = \frac{4}{\hbar} \langle n\rangle F\left(\frac{\tparam}{k_B T}\right) \frac{\uparam^2}{\tparam},
\end{equation}
where  $\langle n\rangle  = N (\mass \omega^2 \lspacing^2/4\pi k_B T)^{3/2}$ is the filling-weighted average lattice filling, $k_B$ is Boltzmann's constant, $N$ is the atom number, $T$ is the temperature at equilibrium, and $F(\tparam/k_B T) \approx 0.3$ is a numerical factor that does not depend strongly on $\tparam/k_B T$ in the regime we probe \cite{SM}.  Fig.~\ref{fig:tauvss}b compares the prediction from Eq.~\eqref{eq:a} with the measured relaxation times.  We show the normalized relaxation rate $h/t\tau$ versus $U^2/t^2$ with $\langle n\rangle$=0.13, which is the average value across all lattice depths determined by the lattice filling before relaxation. No free parameters were used for the theoretically predicted $\tau$, which is constrained by the known experimental values.

Aside from an apparent saturation at small $U/t$, theory and experiment quantitatively agree.  The source of the disagreement at low $s$ is unknown and may arise from other relaxation processes, such as heating from the lattice light.  Surprisingly, Eq~(\ref{eq:a}) appears to be valid even when $U/t$ is as large as 7.  Despite the fact that the theory is predicated on a quasiparticle picture, we find $h/W\tau\sim4$ at the highest $U/t$, which is a violation of the MIR bound.  We emphasize that there is no law preventing a ``violation"  of the MIR limit---rather, the violation simply indicates that the elementary excitations are overdamped, and it is not sensible to describe them as quasiparticles. It is remarkable that the perturbation theory result can be so accurately extrapolated into this regime.

The extraordinarily fast quasimomentum relaxation we observe is key to enabling in-lattice cooling, which is an outstanding challenge for experiments focused on simulating models of strongly correlated electronic solids \cite{McKay2011}.  Several cooling schemes in optical lattices have been proposed \cite{Zoller2007,Heidrich2009,Bernier2009,Ho2009,Williams2010,Popp2006,Rabl2003,PhysRevLett.115.240402}, but experimental demonstrations have been limited \cite{Bakr2011,PhysRevLett.106.195301}. Notably, the quasimomentum degree of freedom has not been cooled directly and remains hotter than the N\'{e}el temperature in experiments with fermionic atoms \cite{Hart2014}. To demonstrate the possibility of in-lattice quasimomentum cooling, we perform a proof-of-principle experiment using Raman transitions to remove the most energetic atoms from the ground band, allowing the remaining atoms to equilibrate to a lower temperature.  This technique is a momentum-space analog of evaporative cooling \cite{PhysRevA.53.381}.  We repeat the process, thereby showing that this approach can be used as part of an iterative cooling method.

\begin{figure}[htb!]
\includegraphics[width=\columnwidth]{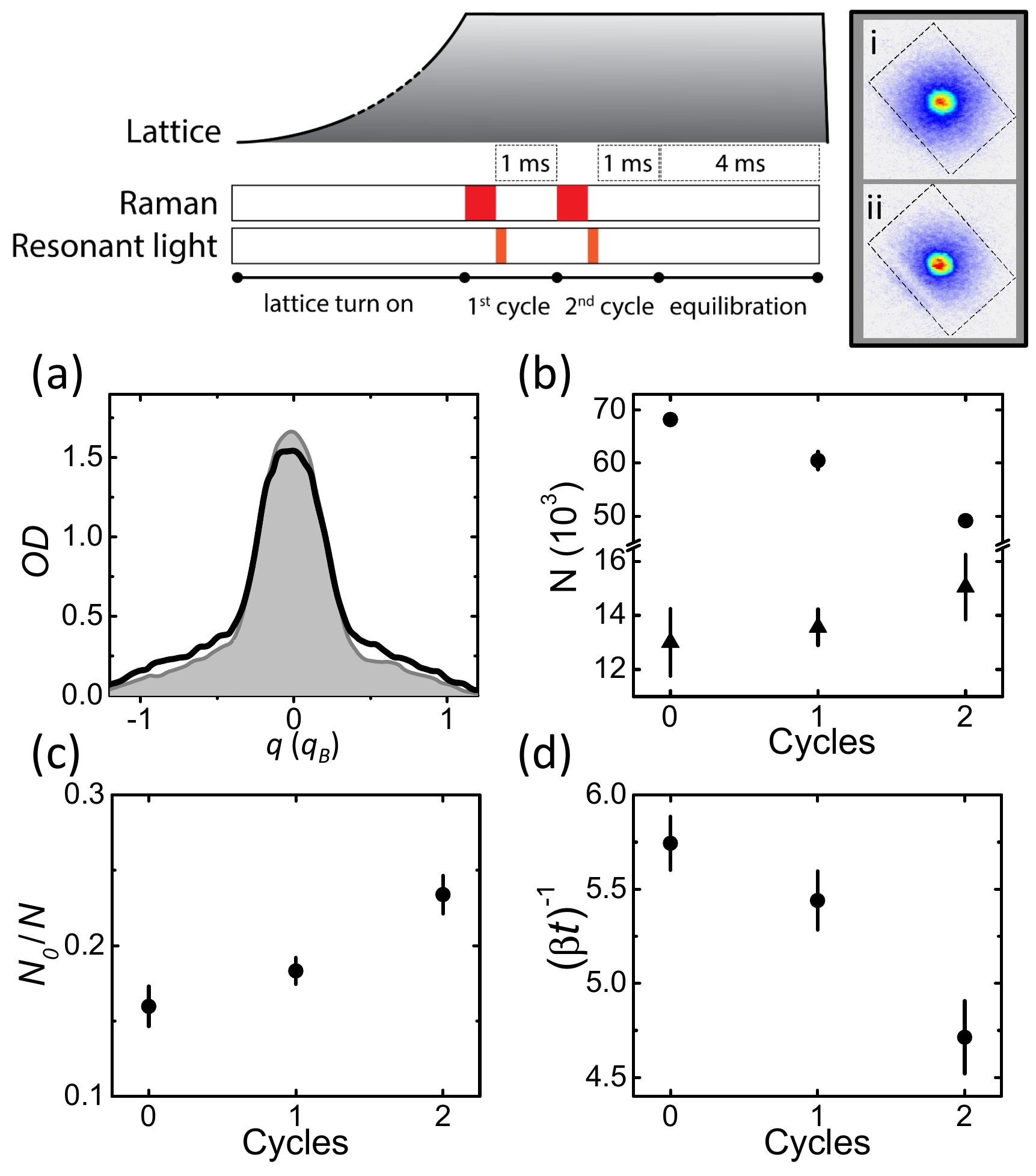}
\caption{\label{fig:coolingPlots} Cooling of the quasimomentum distribution.  The experimental sequence for cooling is shown schematically at the top; the timeline is not to scale.  The TOF images show the quasimomentum distribution immediately before (i) and after (ii) atoms have been removed from the thermal component.  (a) Quasimomentum profile along $k_z$ averaged over $3-4$ images before (black) and after (gray-shaded) two cooling cycles. (b) Atom number in the condensate (\UParrow) and in the thermal component (\CIRCLE) as the cooling cycles are performed. (c),(d) The condensate fraction and the fit parameter $\left(\beta \tparam\right)^{-1}$ (which is monotonically related to the temperature) as a function of the cooling cycles.  Plots (b)-(d) are obtained from fitting the TOF images to a semi-classical model. The error bars represent the standard error of the mean of the measurements averaged for each point.}
\end{figure}

The gas is initially prepared at the same temperature and atom number as for the relaxation experiment in a $s=4 \, E_R$ lattice. For the cooling sequence (depicted in Fig.~\ref{fig:coolingPlots}), the Raman beams and resonant light are subsequently pulsed on for $400\,\mu$s and $50\,\mu$s, respectively. Each Raman pulse is designed to excite only atoms with high quasimomentum, while the condensate is left largely unaffected. Given the spatial configuration of the Raman beams in our experiment, atoms are removed from one side of the FBZ in the $k_z$ direction, as shown in Fig.~\ref{fig:coolingPlots}. At the end of a single cycle, a 1 ms delay is included to allow the quasimomentum distribution to equilibrate. After all the cycles have completed, the gas is held in the lattice potential for 4 ms before bandmapping and imaging.

Figure~\ref{fig:coolingPlots}a shows the quasimomentum profile of the gas along $k_z$ before cooling and after performing two cooling cycles. We observe that the width of the thermal component shrinks and the condensate number grows as the cooling sequence is performed. For a quantitative analysis, we fit the TOF images to the non-interacting, semi-classical model used in the relaxation analysis, with an additional independent Thomas-Fermi profile for the condensate. We use the fit to determine the number of thermal and condensate atoms (and the condensate fraction $N_0/N$) and $\beta t=t/k_B T$.

The results of these fits are shown in Figs~\ref{fig:coolingPlots}b,c, and d.  Figure~\ref{fig:coolingPlots}b shows that the thermal number decreases as atoms are expelled from the trap, while the condensate number increases.  The redistribution of atoms from the thermal component to the condensate is a sign of rethermalization during the cooling cycle. Moreover, the steady increase in the condensate fraction evident in Fig. 4c demonstrates that the entropy per particle is reduced during cooling, and this technique may therefore be used to reach new quantum phases that exist at lower entropy. This conclusion is further reinforced by Fig.~\ref{fig:coolingPlots}d, which shows a monotonic decrease in the temperature inferred from the fits to TOF images. Because bandmapping fails at the edge of the FBZ, a systematic error is made in the fitted $\beta t$.  Nevertheless, the fitted $(\beta t)^{-1}$ is monotonically related to $T$ \cite{McKay2009}, and the decrease in $(\beta t)^{-1}$  signals a reduction in temperature.

A measure of the efficiency for any evaporative cooling scheme is $\alpha=d \log N/d \log T$.  A smaller value of $\alpha$ indicates more efficient cooling---fewer atoms are removed for the same change in $T$.  For our method, $\alpha=1.75\pm0.04$ based on a fit to the data shown in Fig. \ref{fig:coolingPlots}.  This performance compares favorably with recent results for non-lattice gases, including $\alpha\approx1.5$ and 1.9 for ``tilt" evaporation in a hybrid magnetic--optical trap \cite{Chin2008,Lin2009} and $\alpha\approx2.7$ for dipole-trap evaporation of $^{87}$Rb \cite{Olson2013}.

The ultimate limit to the lowest temperature achievable by any cooling method is determined by competition between cooling and heating rates---cooling ceases when the two are equal \cite{McKay2011}.  The heating rate in optical lattices is primarily determined by momentum diffusion resulting from the interaction between the light and atoms \cite{PhysRevA.21.1606}.  This effect can be minimized by detuning the light far from any electronic transition.  In our experiment, the heating rate induced by the lattice light is $0.15\,E_R$/s=25~pK~$k_B$/ms at $s=4$~$E_R$, while the cooling power based on the data shown in Fig. \ref{fig:coolingPlots}d is approximately 9 nK~$k_B$/ms (corresponding to 0.6 $t$/ms).  This extraordinary cooling power is possible because of the high thermalization rate.  In the regime we explore, heating from the lattice is therefore not a limitation to the cooling method.

In principle, more cooling cycles could have been carried out in our experiment.  However, drift in the initial temperature of the gas limited our ability to optimize the pulse sequence. The cooling efficiency can be improved by addressing both sides of the FBZ simultaneously using a small Raman wavevector $\Delta k$ compared with $\pi/d$. Moreover, further enhancement in efficiency can be achieved by adding more pairs of Raman beams to target atoms in the transverse directions to $k_z$ \cite{SM}.  The ultimate cooling limit for this technique, as applied to bosonic atoms, is set by the selectivity in quasimomentum together with the finite momentum spread of the condensate.

In conclusion, we have developed a novel technique for measuring quasimomentum relaxation times in an optical lattice.  We discovered a violation of the MIR bound and showed how our method can be adapted to cool any atomic species. Further studies of relaxation in this system may contribute to our understanding of transport in materials that also display a violation of the MIR bound \cite{RevModPhys.75.1085} and thermalization in closed quantum systems \cite{Olshanni2008}. In the future, the cooling technique could be applied to fermionic atoms trapped in optical lattices in order to reach exotic quantum states that may exist at low entropy per particle.

\acknowledgments{E.M. acknowledges support from ARO grant W911NF-14-1-0003.  B.D., D.C., C.M., and P.R. acknowledge support under grant PHY12-05548 and from the ARO under grant W9112-1-0462.}

\bibliography{CoolingReferences}

\end{document}